\documentclass[10pt,onecolumn,showkeys  ,secnumarabic, nobibnotes, nofootinbib, aps,prd]{revtex4}%
\usepackage[utf8]{inputenc}
\usepackage{amsmath}
\usepackage{amssymb}
\usepackage{amsfonts}
\def\midrule{\hline}\def\bottomrule{\hline}
\usepackage{graphicx}
\usepackage{hyperref}%
\providecommand{\U}[1]{\protect\rule{.1in}{.1in}}

\begin{document}
\title{The Velocity Curves and Three-Dimensional Kinematics\\ of the Milky Way from Gaia DR3}

\author{Petr Zavada}
\author{Karel P\'{\i}\v{s}ka}

\begin{abstract}
We investigate the fundamental kinematic parameters of stellar motions in the
Milky Way using a statistical sample of 27,485,720 stars with measured
three-dimensional velocities from the Gaia DR3 catalogue. From these data, we
determine the local solar velocity $(U_{\odot},V_{\odot},W_{\odot})  =  \left(
10.5\pm0.7,14.5\pm0.9,7.33\pm0.1\right) ~\mathrm{km/s}$, the average rotation
velocity\linebreak   $\mathrm{V}_{0} = \left(  228\pm4\right) $ km/s, and the circular velocity $\mathrm{V}_{c} = \left(  234\pm4\right) $ {km/s,}
 both at the radius
$R_{\odot}$. Next, within the range of $3-14$ kpc, we derive the Galactic
circular velocity curve $\mathrm{V}_{c}(R)$ and determine the average spherical components of the
stellar velocities as functions of both Galactocentric radius $R$ and distance
from the Galactic plane $\left\vert Z\right\vert $. Two-dimensional $R-Z$ maps
of these components reveal a complex three-dimensional kinematic structure. In
particular, the pattern observed at $R\lesssim5$ kpc can be associated with the
influence of the Galactic bar. To facilitate the description of the
observations, we employ a simplified axisymmetric Monte Carlo model with
several free parameters that reproduces the main kinematic trends.
\end{abstract}

\keywords{Milky Way; galactic kinematics; statistical methods; Monte Carlo simulation}

\affiliation{Institute of Physics of the Czech Academy of Sciences, Na Slovance 2, 182 21
Prague 8, Czech Republic}
\email{zavada@fzu.cz}
\accepted{for publication in Galaxies}
\maketitle


\section{Introduction}

\label{introduction}The Milky Way (MW) represents a unique
laboratory for studying gravity and for understanding the formation and
evolution of galaxies. In~recent years, the~Gaia space observatory has
provided an unprecedented amount of highly precise astrometric, photometric,
and spectroscopic data for millions of stars in the MW~\cite{ga3}. The~scientific exploitation of these data has led to thousands of studies devoted
to Galactic structure, stellar populations, and~Galactic~dynamics.

In particular, the~Gaia DR3 catalogue provides exceptionally rich input data
for constructing a detailed kinematic map of the MW, far beyond what was
previously achievable. These data reveal structures on a broad range of
spatial and dynamical scales, from~gravitationally bound binaries and multiple
stellar systems~\cite{za2,za3} to streaming motions associated with spiral
arms~\cite{ant16,ant18}, local velocity fluctuations and turbulence, and~the
global rotation of the Galactic disk and~halo.

The observed Galactic rotation strongly suggests the presence of dark matter,
which is believed to constitute a substantial fraction of the total mass of
the MW (see
~\cite{bha14,eil19,mro19,rei14,bur13,chr20,hua16,jia21,weg19,abl17,beo, ou} and
the overview in~\cite{sof20}). Recent analyses of MW kinematics have provided
increasingly accurate determinations of the Galactic circular velocity curve, defined
as the radial dependence of the galaxy circular velocity~\cite{beo,ou,wan22,lab,zho,jia}. Circular velocity curves
represent one of the principal observational constraints for models of
Galactic mass distribution and dark matter~halos.

In addition to gravity, the~formation and evolution of stars are governed by
the forces of the microworld, namely the strong and electroweak interactions
described by the Standard Model of particle physics. However, despite
extensive observational and theoretical efforts, the~microscopic nature and
origin of dark matter remain~unresolved.

Another important area of current research concerns the detailed mapping of
kinematic substructures and deviations from axial symmetry within the MW,
including spiral-arm perturbations, resonances induced by the Galactic bar,
phase-space streams, and~vertical oscillatory motions
~\cite{gaia2023,ga1,kaw18,lop19,ram18,wan22,wan20,kus17,kaw19,tia17}. A~comprehensive overview of the integrated, structural, and~kinematic properties
of the Galaxy is provided in~\cite{bla16}.

In general, the~kinematic structure of galaxies also provides important clues to their evolutionary history. Observations have revealed galaxies with unusual kinematic properties, suggesting that their disks may have undergone significant changes over time. Such evolutionary processes could result in an atypical orientation of the galactic disk with respect to the dark matter halo. Previous studies have generally indicated an alignment between the angular momentum of dark matter and that of the baryonic component of galaxies~\cite{nav97,steph15,codis18,okabe18,steph25}. Departures from this general picture may arise from interactions with smaller galaxies, including mergers or accretion events involving a more massive host galaxy. The~MW itself may have experienced such interactions, and~certain perturbations in its kinematic structure could provide observational evidence of these past~events.

The primary objective of this paper is to derive the circular velocity curves of the MW
using data from the Gaia observatory, specifically a subset of sources with
measured radial velocities from the DR3 catalogue~\cite{kat}. These data enable
a complete 3D reconstruction of stellar kinematics in arbitrary coordinate
systems. Our analysis focuses on the distribution of stellar velocity components
in spherical coordinates and on their dependence on spatial position. This
work extends our previous 2D kinematic analysis based on the full DR3 catalogue,
in which sources lacking radial velocity measurements constitute the dominant
fraction of the dataset~\cite{za4}.

Section~\ref{definitions} introduces the basic concepts and variables used
throughout the study. Section~\ref{data} describes the criteria used for
selecting stars and defines the angular sectors of the Galactic system that
are suitable for separate analysis. A~useful tool for modelling derived
velocity distributions is a simple Monte Carlo (MC) simulation with several
free parameters, the~algorithm for which is described in Section~\ref{simul}.
Essentially, the~same model was employed in our earlier 2D analysis; however,
the present 3D approach enables a more precise determination of the free~parameters.

The results are presented in Section~\ref{results}. In~addition to circular velocity
curves, our analysis also provides more general velocity maps in $R-Z$
coordinates. The~accuracy of these results depends crucially on the precise
determination of the Sun's motion, which coincides with the
motion of the Gaia observatory. The~independent calculation of the
Sun's local velocity and the Galaxy's mean
rotational velocity at the radius of the Sun's orbit is an
important part of our analysis. In~Section~\ref{conclusion}, the~results
obtained are discussed and, where appropriate, compared with the results of
previous~studies.

\section{Definition of~Variables}

\label{definitions}The positions of sources in Gaia data are expressed in
galactic angular coordinates: longitude ($l$) and latitude ($b$). Using
parallax, we can also define the distance ($r)$ of the source from the Sun.
The proper motion of the stars in Gaia data is represented by the~vector
\begin{equation}
{\mu}_{ICRS}=\left(  \mu_{\alpha}^{\ast},\mu_{\delta}\right)  ;\qquad
\mu_{\alpha}^{\ast}\equiv\mu_{\alpha}\cos\delta,\label{dv1}%
\end{equation}
whose components are angular velocities in the directions of right ascension
and declination in the ICRS. For~our analysis, we prefer the
representation of proper motion in the galactic reference frame:
\begin{equation}
{\mu}_{gal}=\left(  \mu_{l}^{\ast},\mu_{b}\right)  ,\qquad\mu_{l}^{\ast}%
\equiv\mu_{l}\cos b,\label{dv2}%
\end{equation}
which is obtained by the transformation given in~\cite{ga4}. Then, the~transverse 2D {velocity} 
 $\mathbf{v}_{T}$ defined as%
\begin{equation}
\mathbf{v}_{T}=\left(  v_{l},v_{b}\right)  [\mathrm{km/s}]\mathbf{=}%
4.7406\,\,r[\mathrm{kpc}]\,\left(  \mu_{l}^{\ast},\mu_{b}\right)
[\mathrm{mas/year}],\label{dv3}%
\end{equation}
where $v_{l},v_{b}$ are velocity components\ in directions of increasing
latitude $\left(  l\right)  $ and longitude $\left(  b\right)$, and the
distance $r$ is obtained from the parallax $p$:
\begin{equation}
r[\mathrm{kpc}]=\frac{1}{p[\mathrm{mas}]-\mathrm{ZP}}\label{dv4}%
\end{equation}
where ZP is the value of the parallax zero-point for Gaia EDR3 and Gaia DR3
returned by the Python \texttt{gaiadr3\_zeropoint} package~\cite{pramos2025}
based on the functions described in~\cite{ga2}. We work with data that
also include radial velocity ($v_{r}$), so we have complete 3D data on
position and velocity relative to the Sun's position.
For our analysis, it is convenient to express spherical components
$v_{l},v_{b},v_{r}$ in the local standard of rest (LSR) frame. In~fact, there are two LSR approaches:

\begin{enumerate}
\item[(i)]Kinematic---based on the average motion of nearby stars. The~key
parameter is $V_{0}$, the~average velocity of the stars along the circular path of radius
$R_{\odot}$.
\item[(ii)]{Dynamic}
---a hypothetical point that would orbit the Galaxy along
a perfectly circular path at a given distance from the Galactic centre. The~key parameter is $V_{c}$, the~circular velocity of this point along the path of radius
$R_{\odot}$. 

\end{enumerate}

In the first step, the~analysis is conducted within a kinematic framework. Subsequently, in~Section~\ref{results}, selected results are appropriately corrected to account for the effects associated with the dynamic~formulation. 

If $\Delta\mathbf{V}_{\odot}$ is the Sun's velocity relative to the average
velocity of the surrounding stars, then the corresponding LSR components take
the form%
\begin{equation}
w_{\alpha}=v_{\alpha}+\Delta\mathbf{V}_{\odot}\mathbf{.n}_{\alpha}\label{dv5}%
\end{equation}
and satisfy%
\begin{equation}
\left\langle \mathbf{w}\right\rangle \rightarrow\left(  \mathrm{0}%
,\mathrm{0},\mathrm{0}\right)  \text{ for }r\rightarrow0\label{dv11}%
\end{equation}
independently in various sectors around the Sun. The~practical application of
this definition is demonstrated in Section~\ref{results}. The~resulting
$\Delta\mathbf{V}_{\odot}$ may, of~course, depend to some extent on the choice
of these sectors. This inevitable uncertainty must be included in the
measurement error. The~vectors $\mathbf{n}_{\alpha}$ define directions of
increasing coordinates $\alpha=\ l,b,r$. In~terms of Cartesian coordinates, we
have%
\begin{equation}
\mathbf{w=}\left(  \mathrm{w}_{x},\mathrm{w}_{y},\mathrm{w}_{z}\right)
=w_{l}\mathbf{n}_{l}+w_{b}\mathbf{n}_{b}+w_{r}\mathbf{n}_{r}.\label{dv5a}%
\end{equation}

In the next step, we transform spherical galactic LSR coordinates and
velocities to the Galactocentric representation:%
\begin{equation}
l,b,r,w_{l},w_{b},w_{r}\rightarrow\Phi,\Theta,R,V_{\Phi},V_{\Theta}%
,V_{R},\label{dv6}%
\end{equation}
where
\begin{equation}
V_{\alpha}=\mathbf{w.N}_{\alpha};\qquad\alpha=\ \Phi,\Theta,R.\label{dv7}%
\end{equation}

The explicit forms of the unit vectors $\mathbf{n}_{\alpha}$ and
$\mathbf{N}_{\alpha}$ are given in {Appendix} 
 \ref{app11}, where the relationship
between galactic and Galactocentric coordinates is defined. For~clarity, we
use the following convention throughout the paper: positions, velocities, and~their coordinates are denoted in lowercase (uppercase) in the galactic
(Galactocentric) reference frame. The~LSR frame rotates around the Galactic
centre, so the total Galactocentric\linebreak   velocity reads%
\begin{equation}
\mathbf{W}\left(  \mathbf{R}\right)  =\mathbf{w}\left(  \mathbf{R}\right)
+\mathbf{V}_{LSR};\qquad W_{\alpha}=\mathbf{W.N}_{\alpha},\qquad
\mathbf{V}_{LSR}=\left(  0,\mathrm{V}_{0},0\right)  ,\label{dv8}%
\end{equation}
where velocity $\mathrm{V}_{0}>0$ is defined as the average rotation velocity
of stars in the Galactic plane and at the radius of the Sun's orbit $R_{\odot}$.
For our analysis, we use the value \mbox{$R_{\odot}\approx8.178$ kpc~\cite{gra19}.}
The rotation velocity of a star reads
\begin{equation}
W_{\Phi}={V}_{\Phi}-\mathrm V_{0}.\label{dv10}%
\end{equation}

The direction of increasing $\Phi$ (or $\mathbf{N}_{\Phi}$) is opposite to the
direction of the Galaxy's rotation, which explains the negative value of
$W_{\Phi},$\ see Figure~\ref{a1} in~{Appendix}
 \ref{app11}.

The velocities $V_{\Phi},W_{\Phi},V_{\Theta},V_{R}$\ are the subject of
statistical analysis. The~function
\begin{equation}
\mathrm{V}(R)=-\left\langle W_{\Phi}\left(  \mathbf{R}\right)  \right\rangle
\label{dv12}%
\end{equation}
averaged in the galactic plane at fixed $R$ is the input for the calculation of the 
circular velocity curve $\mathrm{V}_{c}(R)$. Velocity $\mathrm{V}_{0}$\ defined above is a special
case: $\mathrm{V}_{0}=$ $\mathrm{V}(R_{\odot}).$ The rotation curve $\mathrm{V}(R)$ reflects the collective
rotation velocity rather than the velocity of a test particle
~\cite{eil19}. We also study the functions $\mathrm{V}(R_{c}%
,Z)=-\left\langle W_{\Phi}\left(  \mathbf{R}\right)  \right\rangle $ and
$\left\langle V_{R}(R_{c},Z)\right\rangle $ depending on two cylindrical
coordinates (\ref{me2a}). The~average $\left\langle ...\right\rangle $ denotes
the average value in the corresponding bin centred at the point $(R_{c},Z)$.

\section{Dataset}

\label{data}
The data sectors of the sky used for analysis are defined in Table \ref{Tsectors}. 

\begin{table}[]%
\small
\caption{{Sectors} 
 of the sky and the number of stars in them. {The sectors} N$_{0}$, Q$_{0}$ and S$_{0}$ cover the entire range of \textbf{\textit{l}}: \boldmath{$\left\langle 0,360\right\rangle $}.}
\label{Tsectors}%


\begin{tabular}{ccccc}
\toprule
\textbf{\textit{b}[deg]}${\diagdown}$\textbf{\textit{l}[deg]} & \boldmath{$\left\langle
-45,45\right\rangle $} & \boldmath{$\left\langle 45,135\right\rangle $} & \boldmath{$\left\langle
135,225\right\rangle $} & \boldmath{$\left\langle 225,315\right\rangle $}\\\midrule
$\left\langle 20,90\right\rangle $ & \multicolumn{4}{c}{N$_{0}$:
3,743,420}\\\midrule
$\left\langle 5,20\right\rangle $ & N$_{1}$: 1,779,692 & N$_{2}$: 1,445,772 &
N$_{3}$: 1,003,565 & N$_{4}$: 1,431,995\\\midrule
$\left\langle -5,5\right\rangle $ & \multicolumn{4}{c}{Q$_{0}$:
7,300,540}\\\midrule
$\left\langle -5,5\right\rangle $ & Q$_{1}$: 2,589,645 & Q$_{2}$: 1,705,027 &
Q$_{3}$: 973,140 & Q$_{4}$: 2,032,728\\\midrule
$\left\langle -20,-5\right\rangle $ & S$_{1}$: 2,480,036 & S$_{2}$:
1,626,841 & S$_{3}$: 902,638 & S$_{4}$: 1,586,298\\\midrule
$\left\langle -90,-20\right\rangle $ & \multicolumn{4}{c}{S$_{0}$:
4,184,923}\\\bottomrule

\end{tabular}
\end{table}

\begin{figure}[t]
\includegraphics[width=80mm]{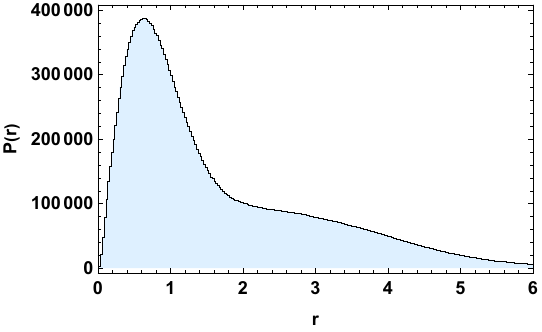}
\includegraphics[width=80mm]{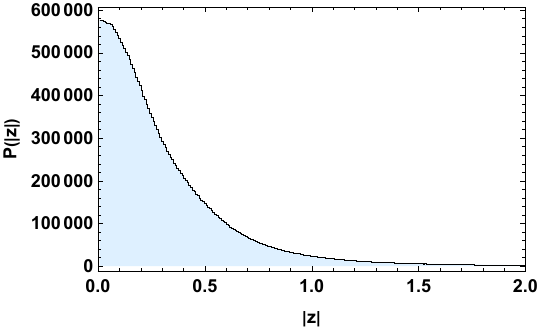}\caption{{Distribution} 
 of distances from the Sun ({\textbf{left}}) and from the galactic plane 
({\textbf{right}}). Units:  
$r,z$ [kpc]. Bin widths: 0.024 and 0.008 kpc.}%
\label{Fdis}%
\end{figure}

\begin{figure}[t]
\includegraphics[width=160mm]{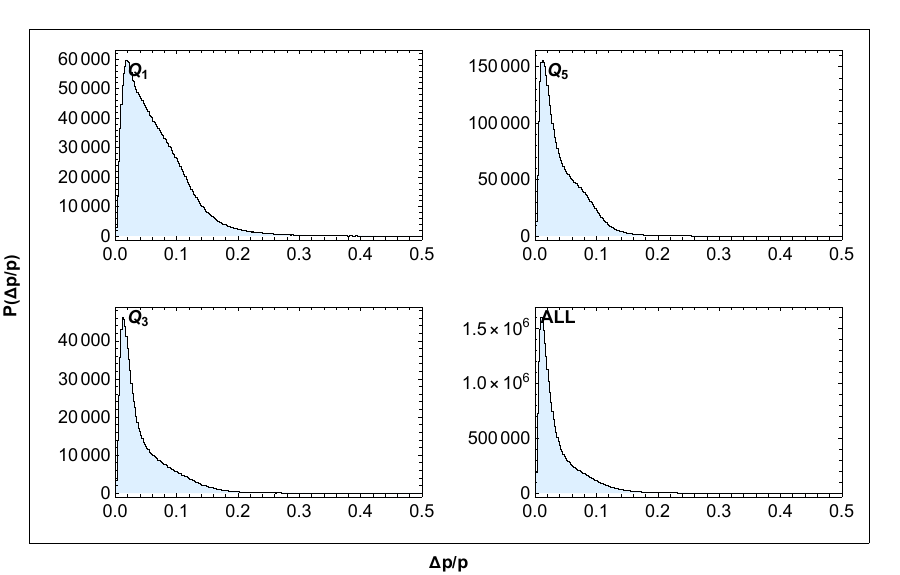} \caption{{Distribution}
 of relative parallax errors across different sectors. Bin width: 0.002.}%
\label{Fpar}%
\end{figure}

From the set of all Gaia DR3 sources with a defined radial velocity
(33,812,183 stars)~\cite{kat}, we only accept sources with a positive parallax
$p$ and limited distance $r\leq6$ kpc that satisfy the following
condition:\vspace{-6pt}%
\begin{equation}
\Delta v_{T}\leq15\text{ km/s},\qquad\Delta v_{r}\leq10\text{ km/s}%
,\label{ds1}%
\end{equation}
where we use the relation~\cite{za4}
\begin{equation}
\frac{\Delta v_{T}}{v_{T}}\approx\sqrt{\left(  \frac{\Delta p}{p}\right)
^{2}+\frac{\left(  \mu_{\alpha}^{\ast}\Delta\mu_{\alpha}^{\ast}\right)
^{2}+\left(  \mu_{\delta}\Delta\mu_{\delta}\right)  ^{2}}{\mu^{4}}};\qquad
\mu=\sqrt{\mu_{\alpha}^{\ast2}+\mu_{\delta}^{2}}.\label{ds2}%
\end{equation}
All parameters, including errors, are included in the Gaia DR3 data. The~resulting star counts are listed in the table. After~these cuts, we are left with a total of 27,485,720 stars for further analysis. The~distributions of
their distances from the Sun and from the galactic plane over all sectors are shown in Figure \ref{Fdis}.%
Note that {the cuts}~(\ref{ds1}) simultaneously select sources with higher
parallax precision. Figure \ref{Fpar} {show}  
 the corresponding distribution of relative parallax errors.
The selected sources meet the condition $\Delta p/p<0.2$ in all sectors, which
is the standard criterion for selecting ``good'' sources. The~first cut on $\Delta v_{T}$ in Equation~(\ref{ds1}) should be regarded as approximate because~covariance terms are neglected in (\ref{ds2}). We have verified that the
narrowing of the cuts does not systematically affect the parameters of the resulting curves in Section~\ref{results}. Most of our calculations focus on
mean values, which means that the resulting errors are much smaller than the
errors of individual entries. The~error bars in the figures in
Section~\ref{results} represent the statistical errors of the average values in the
corresponding bins. These errors are calculated using the formula%
\begin{equation}
\Delta v=\frac{\sigma_{b}}{\sqrt{n_{b}}};\qquad\sigma_{b}=\sqrt{\left\langle
v^{2}\right\rangle -\left\langle v\right\rangle ^{2}},\label{ds3}%
\end{equation}
where $\sigma_{b}$ is the standard deviation of the velocities and $n_{b}$ is
their number in the given bin. \vspace{-6pt}

\section{Algorithm of Monte Carlo~Simulations}

\label{simul} The velocity distributions are compared with a simple MC
probabilistic model proposed in our previous study~\cite{za4}. This model
represents an axially symmetric approximation corresponding to a simplified
scenario in which local velocity fluctuations are averaged out. The~simulated
velocity of a star in the Galactocentric reference frame is given by the~relation
\begin{equation}
\mathbf{W}^{gen}\left(  \mathbf{R}\right)  \mathbf{=}\left(  V_{\Phi
}-\mathrm{V}(R)\right)  \mathbf{N}_{\Phi}+V_{\Theta}\mathbf{N}_{\Theta}%
+V_{R}\mathbf{N}_{R}, \label{ge1}%
\end{equation}
where the velocity $V\left(  R\right)  >0$ is defined \ by Equation~(\ref{dv12}). The~components $V_{\Phi},V_{\Theta}$, and~$V_{R}$ are generated in the local reference frame defined by the basis (\ref{me3a}). The~model is based on the
multinormal distribution
\begin{equation}
P(V_{\Phi}\mathbf{,}V_{\Theta},V_{R};R,Z\mathbf{)\sim}\exp\left(
-\frac{V_{\Phi}^{2}}{2\sigma_{\Phi}^{2}}-\frac{V_{\Theta}^{2}}{2\sigma
_{\Theta}^{2}}-\frac{V_{R}^{2}}{2\sigma_{R}^{2}}\right)  , \label{ge3}%
\end{equation}
with partial asymmetry (different $\sigma_{\Phi}$ for positive and negative
$V_{\Phi}$). The~dependence on $R$ and $Z$ is absorbed in the standard
deviations $\sigma_{\alpha}(R,Z)$. As~a first approximation, we assume
\begin{equation}
\sigma_{\alpha}=\sigma_{\alpha0}+\sigma_{\alpha1}\left\vert Z\right\vert
;\qquad\alpha=\Phi,\Theta,R, \label{r12}%
\end{equation}
where $\sigma_{\alpha0}$ and $\sigma_{\alpha1}$ are free parameters. We have
found that the optimal shape of the distribution $V_{\Phi}$\ is slightly
asymmetric, with different $\sigma_{\Phi}^{\pm}$ for the two opposite
orientations, where $+(-)$ denotes {in (against)} 
 the direction of rotation:%
\begin{equation}
\sigma_{\Phi}^{+}=\sigma_{\Phi0},\qquad\sigma_{\Phi}^{-}=\sigma_{\Phi0}%
+\sigma_{\Phi1}^{-}\left\vert Z\right\vert . \label{r13c}%
\end{equation}

Thus, only $\sigma_{\Phi}^{-}$ depends on $\left\vert Z\right\vert $. This
asymmetry also reflects the effective deceleration $\left\langle \Delta
V_{\Phi}\right\rangle $ of the collective rotation for larger $\left\vert
Z\right\vert $:
\begin{equation}
-\left\langle W_{\Phi}\left(  \left\vert Z\right\vert \right)  \right\rangle
=\mathrm{V}_{0}-\left\langle \Delta V_{\Phi}\right\rangle \approx
\mathrm{V}_{0}-\varkappa\left\vert Z\right\vert ;\qquad\varkappa=\sqrt
{\frac{2}{\pi}}\sigma_{\Phi1}^{-}. \label{mc1}%
\end{equation}

\begin{figure}[t]
\includegraphics[width=160mm]{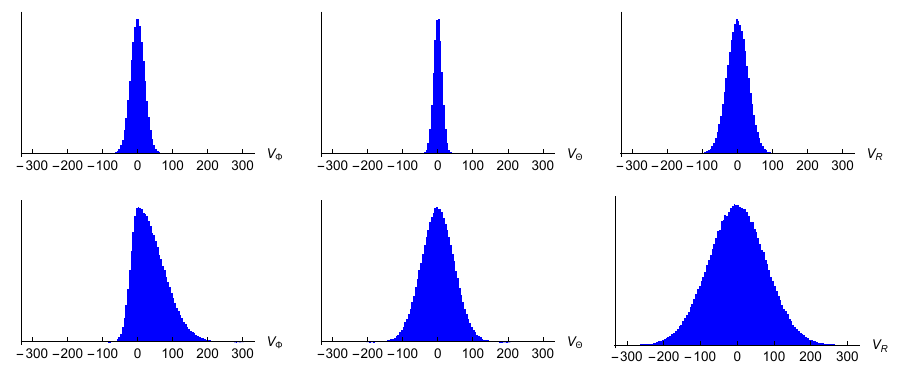}\caption{Distribution of simulated velocities $V_{\Phi},V_{\Theta}$, and~$V_{R}$ at $Z=0$ ({\textbf{{top} 
 row}}) and $Z=1.5$ kpc ({\textbf{bottom row}}). Units: $V_{\alpha}$ [km/s]. }%
\label{FMC}%
\end{figure}

It should be noted that the parameter $\varkappa$ $\left(  \text{or }\sigma
_{\Phi1}^{-}\right)$ represents only an average over the sectors $N_{\alpha
}$ and $S_{\alpha}$. In~Section~\ref{results}, we show that the actual
situation is somewhat more complex. The~Wolfram Mathematica code of the
generator is available {at} 
\url{https://www.fzu.cz/~piska/Catalogue/genkinJUN26.nb } (accessed on 8 October 2026). Figure~\ref{FMC}
 {shows}
 examples of the distribution of simulated velocities
$V_{\Phi}\mathbf{,}V_{\Theta}$ and $V_{R}$. The~distributions $V_{\Theta}$ and
$V_{R}$ are symmetric for any $\left\vert Z\right\vert $, whereas the distribution
$V_{\Phi}$ is asymmetric for $\left\vert Z\right\vert >0$.

The simulated distributions at different positions are obtained as follows. The~position $\left(  l,b,r\right)  $ of the source in the galactic reference frame is transformed using Equation (\ref{me1}) into the Galactocentric spherical
coordinate system $\left(  \Phi,\Theta,R\right)$. Then, at this position, the~velocity $\mathbf{W}^{gen}\left(  \mathbf{R}\right)  $ is generated according to the distribution (\ref{ge1}) using the parameters from Table~\ref{Tab2}. The~optimal parameters $\sigma_{\alpha}$ listed in the table are obtained through comparison with the real data presented in the following section. The~generated velocity can be transformed back to the galactic reference frame:%
\begin{equation}
w_{\alpha}^{gen}\mathbf{(r)}=\left(  \mathbf{W}^{gen}\left(  \mathbf{R}%
\right)  -\mathbf{V}_{LSR}\right)  \mathbf{.n}_{\alpha};\qquad\alpha=l,b,r.
\label{mc2}%
\end{equation}

In this way, for~any source defined by input $\left(  l,b,r,w_{l},w_{b}%
,w_{r}\right)$ or $\left(  \Phi,\Theta,R,V_{\Phi}%
\mathbf{,}V_{\Theta},V_{R}\right)$, we generate the vector $\left(
l,b,r,w_{l}^{gen},w_{b}^{gen},w_{r}^{gen}\right)  $ or $\left(  \Phi
,\Theta,R,V_{\Phi}^{gen}\mathbf{,}V_{\Theta}^{gen},V_{R}^{gen}\right)  $.  
To compare the model simulations with the
measured data, we also need to know the local velocity of the Sun 
$\Delta\mathbf{V}_{\odot}$ and~the rotation velocity $\mathrm{V}_{0}$, which
we also determine in the following~section.

It goes without saying that different methods used to analyse different data
samples can lead to slightly different parameters. Since the 3D analysis is
more consistent, we give preference to the parameters $\sigma_{\alpha
},\mathrm{V}_{0}$, and~$\Delta\mathbf{V}_{\odot}$ obtained in this~work.

\begin{table}[]\small
\caption{{Optimal} 
 parameter values in the MC simulation. Units: $\sigma_{\alpha
0}[\mathrm{km\ s}^{-1}]$, $\sigma_{\alpha1}[\mathrm{km\ s}^{-1}\mathrm{\ kpc}%
^{-1}]$.}%
\label{Tab2}%
\begin{tabular}{ccccccc}
\toprule
\boldmath{$\sigma_{\Theta0}$}~ & \boldmath{$\sigma_{\Phi0}$}~ & \boldmath{$\sigma_{R0}$}~ & \boldmath{$\sigma_{\Theta1}
$}~ & \boldmath{$\sigma_{\Phi1}^{-}$}~ & \boldmath{$\sigma_{R1}$} & \textbf{\boldmath{Ref.}}\\
\midrule
$12$ & $20$ & $30$ & $23$ & $32$ & $32$ & this work\\ 
$18.03\pm0.03$ & $24.35\pm0.04$ & $36.81\pm0.07$ & x & x & x & \cite{ang20}%
\\
$11$ & $20$ & $31$ & x & x & x & \cite{vie22}\\ \bottomrule
\end{tabular}
\end{table} 
\section{Results}
\label{results} 

\begin{figure}[t]
\includegraphics[width=160mm]{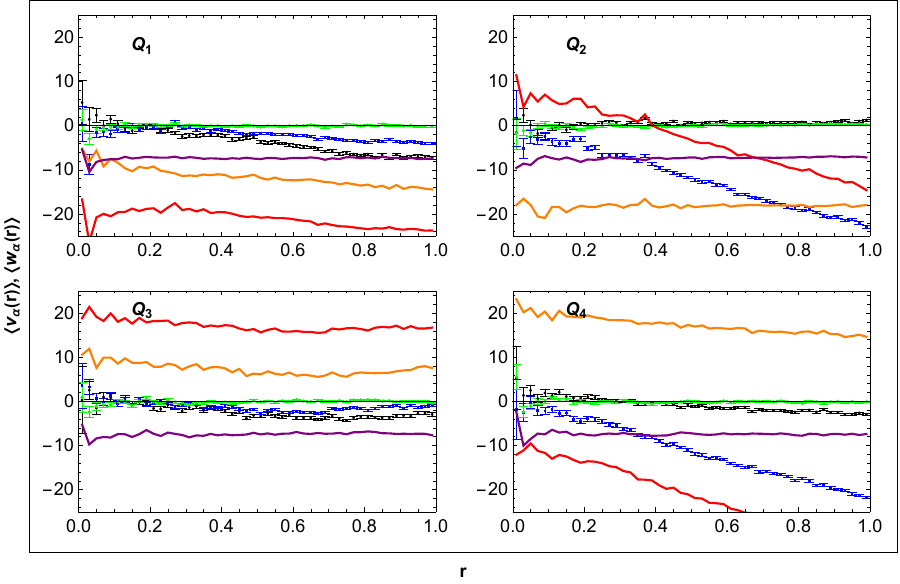}\caption{{Dependence} 
 of mean velocity
($w_{l}-$blue; $w_{b}-$green; $w_{r}-$black; $v_{l}-$red; $v_{b}-%
$purple; and~$v_{r}-$orange) on distance $r$ in sectors $Q_{1}$--$Q_{4}$.
Units: r [kpc], v [km/s].}%
\label{Fga1}%
\end{figure}

In Figure~\ref{Fga1}, {we}
 show the dependence of average galactic velocities on the
distance $r$. We compare the set of galactic velocities related to the Sun
($v_{\alpha}$) with those related to the LSR ($w_{\alpha}$), defined in
Equation~(\ref{dv5}). Once the Sun's local velocity $\Delta\mathbf{V}_{\odot}$ has
been set correctly, the~average velocities $\left\langle w_{\alpha
}\right\rangle $ should satisfy the condition (\ref{dv11}). This condition is
well satisfied for the values used in the figure:
\begin{equation}
\Delta\mathbf{V}_{\odot}=\left(  \Delta V_{x\odot},\Delta V_{y\odot},\Delta
V_{z\odot}\right)  =\left(  10.5\pm0.7,20.5\pm0.9,7.33\pm0.1\right)
\mathrm{km/s}. \label{re4}%
\end{equation}

These values were obtained after a minor correction (through visual fitting) of our previously measured
values in~\cite{za4}, which also includes a comparison with other studies. Since the method for determining $\Delta\mathbf{V}_{\odot}$ is virtually the same in both studies, we adopt the magnitude of systematic errors from the previous study. Statistical errors, on~the other hand, are~insignificant.

\begin{figure}[t]
\includegraphics[width=80mm]{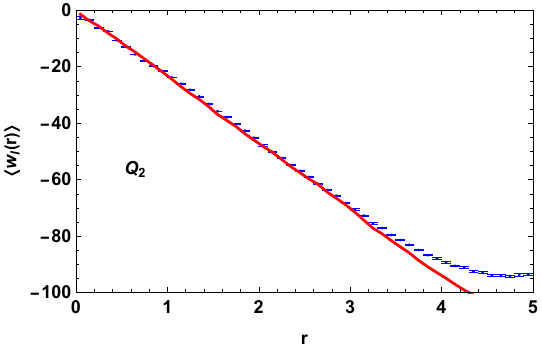}\includegraphics[width=80mm]{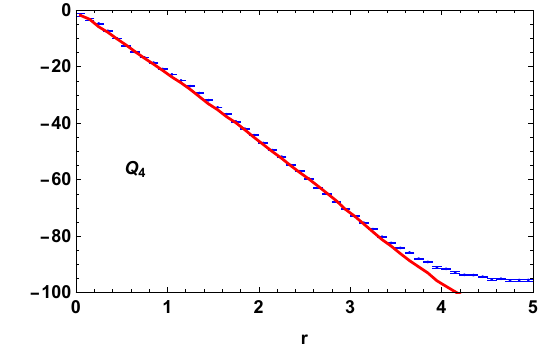}\caption{{The} 
curves $\left\langle w_{l}\right\rangle $ in sectors $Q_{2}$ and $Q_{4}$: data
{(blue)};
simulation model {(red).} Units:\ $r$ [kpc], $w_{l}%
$ [km/s].}%
\label{Fgaw}%
\end{figure}

{The} 
 slope of the curves $\left\langle w_{l}\right\rangle $ in sectors $Q_{2}$
and $Q_{4},$ as we discussed in the cited paper, is a direct consequence
of MW rotation. In~Figure~\ref{Fgaw}, {we}
 show these curves on a larger scale together with their
simulated counterparts given by Equation~(\ref{mc2}). Consistency with the simulation further
confirms the condition (\ref{dv11}) for $\left\langle w_{l}\right\rangle $.
The slope can be explained as follows. In~the plane of the MW disc $\left(
b,z=0\right)  $, the~rotation is defined by the function $\mathrm{V}(R)$ in
(\ref{dv12}), so the average velocity $w_{\alpha}$\ at the position
$\mathbf{r}$ observed from the LSR frame reads%
\begin{equation}
\left\langle w_{\alpha}\mathbf{(r)}\right\rangle =-\left(  \mathrm{V}(R)
\mathbf{N}_{\Phi}\left(  \mathbf{R}\right)  +\mathbf{V}_{LSR}\right)
\mathbf{.n}_{\alpha}\mathbf{(r)};\qquad\alpha=l,b,r. \label{re2}%
\end{equation}

For the curves $\left\langle w_{l}\right\rangle $, we then obtain%
\begin{equation}
\left\langle w_{l}\right\rangle =\mathrm{V}_{0}\left(  \frac{R_{\odot}\cos
l-r}{\sqrt{R_{\odot}^{2}+r^{2}-2R_{\odot}r\cos l}}-\cos l\right)  .
\label{re3}%
\end{equation}

\begin{figure}[t]
\includegraphics[width=80mm]{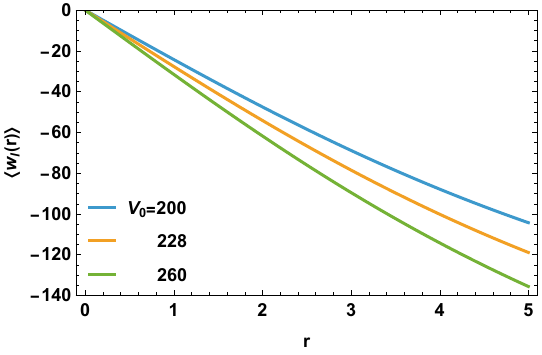}\includegraphics[width=80mm]{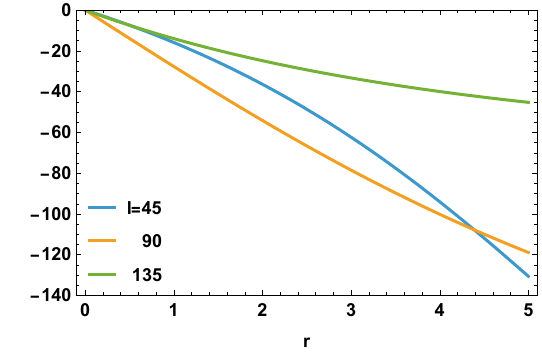}\caption{{Dependence} 
of velocity $\left\langle w_{l}\right\rangle $ on distance $r$ for various
rotation velocities $V_{0}$ (at fixed \mbox{$l=90$ deg)} and various angles $l$ (at
fixed $V_{0}=228$ km/s) in the galactic plane ($b=0$). Units: r [kpc], $w_{l}%
$ [km/s].}%
\label{Fwr}%
\end{figure}

In Figure~\ref{Fwr}, {we} 
 show the set of these curves for different values of the parameters
$\mathrm{V}_{0}$\ and $l$. The~curves in Figure~\ref{Fgaw} correspond to a
superposition of curves over $l$ from the sectors $Q_{2}$ and $Q_{4}$. The MC
simulation is based on the value obtained in the previous study:

\begin{equation}
\mathrm{V}_{0}=\left(  228\pm4\right)  \text{km/s.} \label{re5}%
\end{equation}

It is clear that $\mathrm{V}(R)\approx\mathrm{V}_{0}$ agrees perfectly with
the measured values in the range $r\lesssim3.5$ kpc, which confirms the
previous~result.

For consistent comparison with analyses resulting from the dynamic definition of the LSR (see Section~\ref{definitions}), the~velocities $\Delta
V_{y\odot}$ and $\mathrm{V}_{0}$ require a correction based on the calculation of asymmetric drift
\begin{equation}
\mathrm V_{a}=\mathrm V_{c}-\mathrm V_{0},\label{cor1}%
\end{equation}
where $V_{c}$ is the MW circular velocity at $R_{\odot}$. The~Sun's rotation velocity $V_{\odot\Phi}$ in the Galactocentric frame can be decomposed alternatively as
\begin{equation}
V_{\odot\Phi}=\mathrm V_{0}+\Delta V_{y\odot}=\mathrm V_{c}+\Delta V_{y\odot}^{corr}%
,\label{cor2}%
\end{equation}
so
\begin{equation}
\Delta V_{y\odot}^{corr}=\Delta V_{y\odot}-\mathrm V_{a},\qquad \mathrm V_{c}=\mathrm V_{0}%
+\mathrm V_{a}.\label{cor3}%
\end{equation}

In~\cite{li}, the asymmetric drift $V_{a}$\ around the Sun's position was estimated to be about $6$~km/s. A~similar value can be extracted from {Figure~3}
in \cite{sch10}, where the parameter $\left\langle
v_{R}^{2}\right\rangle $ is replaced with our value of $\sigma_{R0}^{2}=900$ km$^{2}$s$^{-2}$, as shown in Table~\ref{Tab2}. In the standard notation, we can identify
\begin{equation}
U_{\odot}=\Delta V_{x\odot},\qquad V_{\odot}=\Delta V_{y\odot}^{corr},\qquad
W_{\odot}=\Delta V_{z\odot}.\label{cor4}%
\end{equation}

These results are summarised in Table~\ref{Tab3}.\begin{table}[]%

\caption{Sun's local velocity, mean rotation velocity $\mathrm{V}_{0}$, and circular velocity $\mathrm{V}_{c}$. Units: $[\mathrm{km\ s}^{-1}]$.}%
\label{Tab3}%
\begin{tabular}{ccccc}
\toprule
\boldmath{$U_{\odot}$} & \boldmath{$V_{\odot}$} & \boldmath{$W_{\odot}$} & \boldmath{$\mathrm{V}_{0}$} & \boldmath{$\mathrm{V}_{c}$}\\\midrule
$10.5\pm0.7$ & $14.5\pm0.9$ & $7.33\pm0.1$ & $228\pm4$ & $234\pm4$\\
\bottomrule
\end{tabular}

\end{table}

\begin{figure}[t]
\hspace{-0.8em}
\includegraphics[width=140mm]{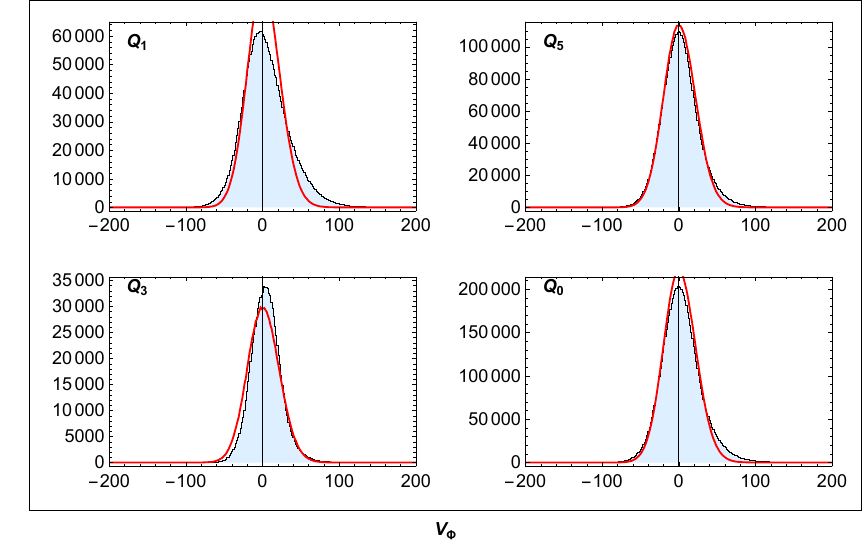} \caption{{Distributions} 
 of $V_{\Phi}$ in sectors $Q_{\alpha}$\ of the galactic plane defined in Table~\ref{Tsectors} and in Equation~(\ref{re1}): data ({light blue}); simulation model ({red}). Unit: $V_{\Phi}$ [km/s]. Bin width: 1.6 km/s.}%
\label{F1FGc}%
\end{figure}

\begin{figure}[t]
\hspace{-0.8em}\includegraphics[width=140mm]{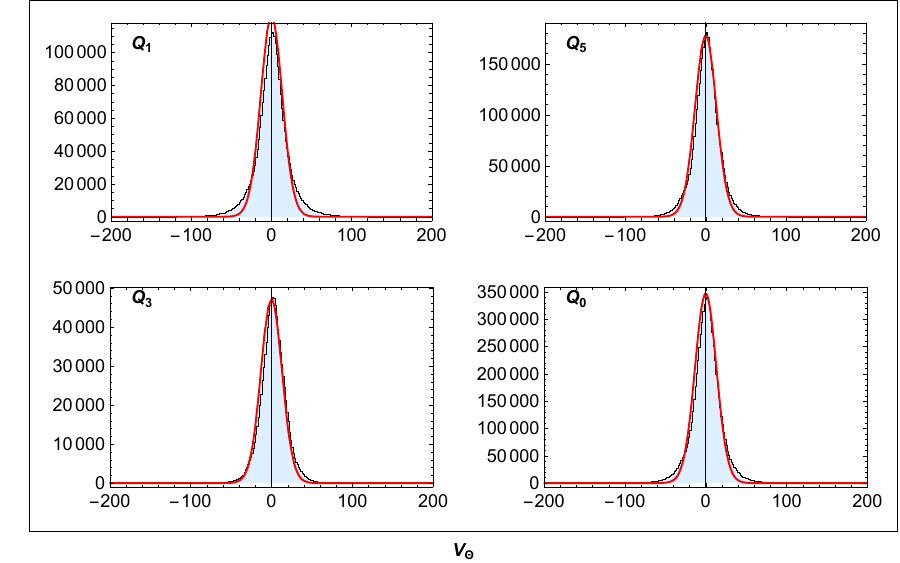} \caption{{{Distributions} 
} of $V_{\Theta}$ in sectors $Q_{\alpha}$\ of the galactic plane defined in Table~\ref{Tsectors} and in \mbox{Equation~(\ref{re1}):} data ({light blue}); simulation model ({red}). Unit: $V_{\Theta}$ [km/s]. Bin width: 1.6 km/s.}%
\label{F1TGc}%
\end{figure}

\begin{figure}[t]
\hspace{-0.8em}\includegraphics[width=140mm]{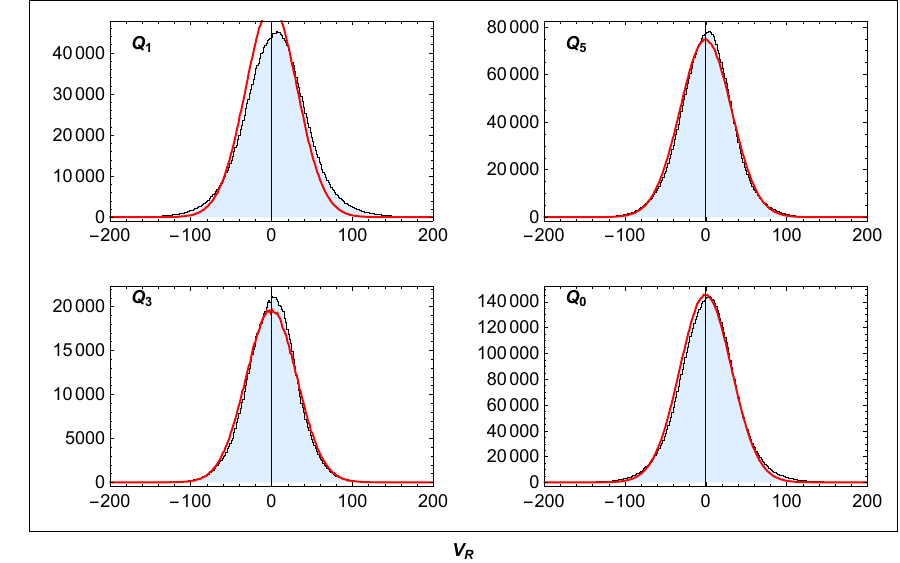}\caption{{{Distributions}
} of $V_{R}$ in sectors $Q_{\alpha}$\ of the galactic plane defined in Table~\ref{Tsectors} and in Equation~(\ref{re1}): data ({light} {blue}); simulation model ({red}). Unit: $V_{R}$ [km/s]. Bin width: 1.6 km/s.}%
\label{F1RGc}%
\end{figure}

The Galactocentric velocities $V_{\Phi}\mathbf{,}V_{\Theta},V_{R}$ are calculated using Equation~(\ref{dv7}). The distributions $P(V_{\Phi}),$ $P(V_{\Theta})$, and $P(V_{R})$ in the sectors $Q_{1}$--$Q_{4}$ covering the galactic plane are shown in Figures~\ref{F1FGc}--\ref{F1RGc}.

\begin{figure}[t]
\hspace{-0.7em}\includegraphics[width=140mm]{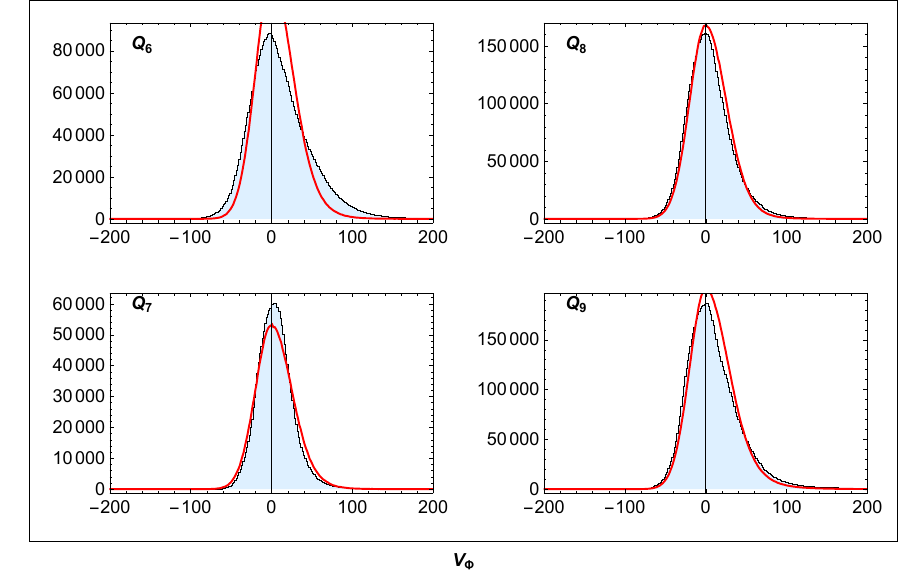}\caption{{{Distributions}
} of $V_{\Phi}$ in sectors $Q_{\alpha}$\ outside the galactic plane defined in Equation~(\ref{re1}):
data ({blue}); simulation model ({red}). Unit:
$V_{\Phi}$ [km/s]. Bin width: 1.6 km/s.}%
\label{F2QQF}%
\end{figure}

\begin{figure}[t]
\hspace{-0.7em}\includegraphics[width=140mm]{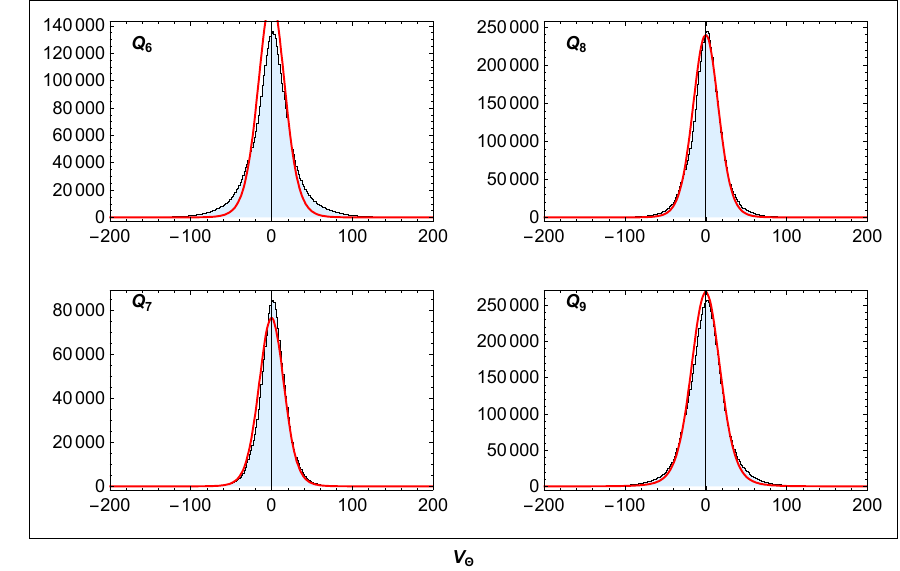}\caption{{{Distributions}
} of
$V_{\Theta}$ in sectors $Q_{\alpha}$\ outside the galactic plane defined in
Equation~(\ref{re1}): data ({light blue}) and the simulation model
({red}). Unit: $V_{\Theta}$ [km/s]. Bin width: 1.6 km/s.}%
\label{F2QQT}%
\end{figure}

\begin{figure}[t]
\hspace{-0.7em}\includegraphics[width=140mm]{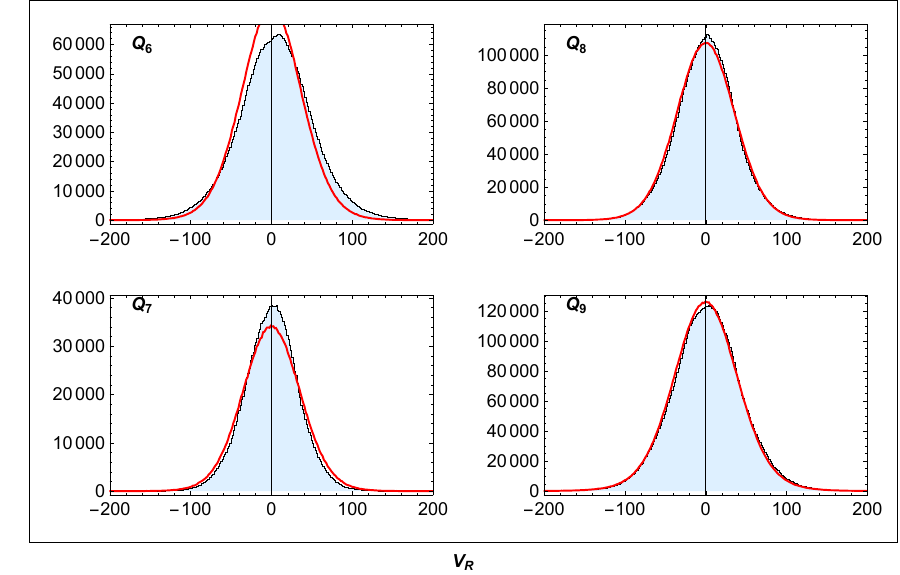}\caption{{{Distributions} 
} of $V_{R}$ in
sectors $Q_{\alpha}$\ outside the galactic plane defined in Equation (\ref{re1}): data
({light blue}); simulation model ({red}). Unit: $V_{R}%
$ [km/s]. Bin width: 1.6 km/s.}%
\label{F2QQR}%
\end{figure}

The same distributions for other sectors (outside the galactic plane) are shown in Figures~\ref{F2QQF}--\ref{F2QQR}. In~the figures, we merge the sectors, which, due to rotational symmetry, give very similar results:%
\begin{align}
&  Q_{5}=Q_{2}+Q_{4}\qquad Q_{6}=S_{1}+N_{1}\qquad Q_{7}=S_{3}+N_{3}%
\label{re1}\\
&  Q_{8}=S_{2}+N_{2}+S_{4}+N_{4}\qquad Q_{9}=S_{0}+N_{0}\nonumber
\end{align}

Figures~\ref{F1FGc}--\ref{F2QQR} are used to determine the parameters of the simulated curves in Table~\ref{Tab2}.

Since these parameters generate 24 different images, we optimised them through visual fitting.  The~distributions in Figures~\ref{F1FGc}--\ref{F1RGc} correspond to the galactic plane ($Z=0$). Their simulated shapes are controlled solely by the parameters $\sigma_{\alpha0}$, which are fixed based on these figures. The~remaining parameters $\sigma_{\alpha1}$ are then fixed so that the simulation best matches the data in Figures~\ref{F2QQF}--\ref{F2QQR}, which correspond to the sectors outside the galactic plane ($|Z|>0$). In~all figures except~for sectors $Q_{1}$ and $Q_{6}$ oriented towards the MW centre ($-45^{0}<l<45^{0}$), we observe very good agreement between the MC simulation and the data. A~slight asymmetry of the $V_{\Phi}%
$\ distribution (both data and simulation) is observed outside the galactic plane in Figure~\ref{F2QQF}. A~slight widening with increasing $\left\vert
Z\right\vert $ is most clearly evident in Figures~\ref{F1RGc} and \ref{F2QQR} when comparing the $V_{R}$ distributions in sectors $Q_{0}$ and $Q_{9}$. All parameters are close to the values we obtained previously~\cite{za4} and in other studies listed in the table. In~the cited studies, the~velocity dispersions are related to the cylindrical coordinate system, whereas ours are related to the spherical one. However, since small $\left\vert Z\right\vert $ values dominate the data, it can be assumed that $V_{\Theta}\approx
V_{Z},V_{R}\approx V_{R_{c}}$ and $\sigma_{\Theta0}\approx\sigma_{Z}%
,\sigma_{R0}\approx\sigma_{R_{c}}$ correspondingly.   

\begin{figure}[t]
\includegraphics[width=80mm]{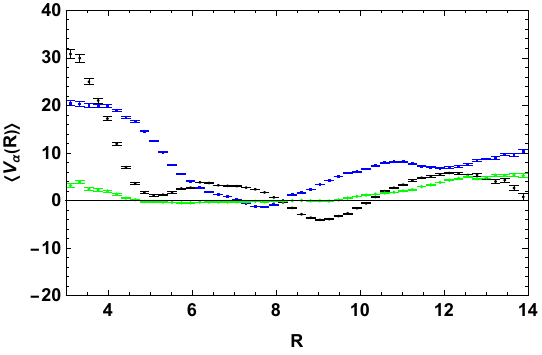}\includegraphics[width=80mm]{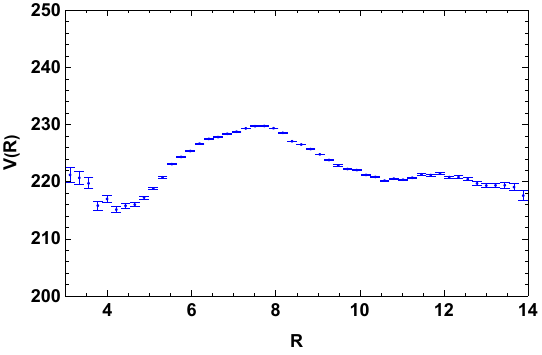}\caption{
Dependence of $\left\langle V_{\Phi}\right\rangle $ ({blue})$,\left\langle
V_{\Theta}\right\rangle $ ({green}), and $\left\langle V_{R}%
\right\rangle $({black}) on radius $R$ in the galactic plane
(left) in sector Q$_{0}$ and rotation curve V($R$) (right) in the same sector but with an additional cut $\left\vert
b\right\vert <2$ deg. Units: V [km/s], $R$ [kpc].}
\label{FGv4}%
\end{figure}

In the left panel of Figure~\ref{FGv4}, {we} 
 show the velocity curves $\left\langle V_{\Phi}\right\rangle
,\left\langle V_{\Theta}\right\rangle $, and $\left\langle V_{R}\right\rangle $
in sectors covering the galactic plane depending on the radius $R$ in this
plane. All curves give zero at the point $R_{\odot}$, which once again
confirms the correct setting of $\Delta\mathbf{V}_{\odot}$ in our calculation.
The curves fluctuate around zero, while the simulations yield a net zero (not
shown in the figure). Local kinematic substructures generate deviations $\Delta
v\lesssim5$ km/s. A~more pronounced increase in $\left\langle V_{\Phi
}\right\rangle $ (which means a decrease in rotation velocity) is evident for
$R\lesssim5$ kpc. The~apparent increase in $\left\langle V_{R}\right\rangle $
for $R\lesssim4$ kpc may indicate that a slight centrifugal component
accompanies the lower rotation velocity in this region. The~resulting rotation velocity 
curve $\mathrm{V}(R)=\mathrm{V}_{0}-\left\langle V_{\Phi}\right\rangle $ is
shown in the right panel, where we used an additional cut $b<2$ deg (see note
in the following section). The~high number of stars in sector $Q_{0}%
$\ makes it possible to obtain curves with very good resolution of local
kinematic~substructures.

Table~\ref{Tab4}  {contains} 
 the corresponding circular velocity curve%
\begin{equation}
\mathrm{V}_{c}\left( R\right) =\mathrm{V}\left( R\right) +\mathrm{V}_{a},
\label{re6}
\end{equation}%
where, in~accordance with~\cite{li}, the asymmetric drift in the galactic plane $Z\approx 0$ is assumed to be $\mathrm{V}_{a}\approx 6$ km/s regardless of radius $R$.

\begin{table}[]

\renewcommand{\arraystretch}{1.2}
\newcommand{\aboverulesep}{.1pt}
\newcommand{\belowrulesep}{.1pt}
\caption{{Circular} 
 velocity curve $\mathrm{V}_{c}\left( R\right)$ with corresponding errors in the defined $\Delta R$ bin. N is the number of entries in a given bin. Units: $\mathrm{V}_{c}$ [km/s], $R$ [kpc].}
\label{Tab4}
\begin{tabular}{r|rrrrrrrrrrr}
\toprule 
\boldmath{$R$} & \textbf{3.1} & \textbf{3.3} & \textbf{3.5} & \textbf{3.7} & \textbf{3.9} & \textbf{4.1} & \textbf{4.3} & \textbf{4.5} & \textbf{4.7} & \textbf{4.9} &\textbf{ 5.1} \\ 
\midrule
$\mathrm{V}_{c}\left( R\right)$ & 227.33 & 228.41 & 224.4 & 223.35 & 222.99 & 221.22 & 222.09 & 221.41
& 222.1 & 223.38 & 224.95 \\ 
$\Delta _{stat}$ & 1.41 & 1.26 & 1.07 & 0.85 & 0.73 & 0.6 & 0.48 & 0.39 & 
0.31 & 0.25 & 0.2 \\ 
$\Delta _{tot}$ & 7.41 & 7.26 & 7.07 & 6.85 & 6.73 & 6.6 & 6.48 & 6.39 & 6.31
& 6.25 & 6.2 \\ 
N & 489 & 647 & 984 & 1411 & 2103 & 3085 & 4872 & 7370 & 11050 & 16730 & 
24148 \\ \midrule
$R$ & 5.3 & 5.5 & 5.7 & 5.9 & 6.1 & 6.3 & 6.5 & 6.7 & 6.9 & 7.1 & 7.3 \\ 
\midrule
$\mathrm{V}_{c}\left( R\right)$ & 226.66 & 228.7 & 230.1 & 230.95 & 232.09 & 233.23 & 233.59 & 233.93
& 234.45 & 234.84 & 235.32 \\ 
$\Delta _{stat}$ & 0.16 & 0.13 & 0.11 & 0.1 & 0.1 & 0.09 & 0.09 & 0.08 & 0.08
& 0.07 & 0.07 \\ 
$\Delta _{tot}$ & 6.16 & 6.13 & 6.11 & 6.1 & 6.1 & 6.09 & 6.09 & 6.08 & 6.08
& 6.07 & 6.07 \\ 
N & 34,811 & 48,698 & 64,079 & 78,064 & 88,319 & 90,507 & 97,609 & 112,446 & 114,642
& 125791 & 146645 \\ \midrule
$R$ & 7.5 & 7.7 & 7.9 & 8.1 & 8.3 & 8.5 & 8.7 & 8.9 & 9.1 & 9.3 & 9.5 \\ 
\midrule
$\mathrm{V}_{c}\left( R\right)$ & 235.68 & 235.73 & 235.4 & 234.96 & 233.57 & 232.63 & 232.2 & 231.43
& 230.6 & 229.59 & 228.78 \\ 
$\Delta _{stat}$ & 0.06 & 0.06 & 0.06 & 0.06 & 0.06 & 0.06 & 0.06 & 0.07 & 
0.08 & 0.09 & 0.1 \\ 
$\Delta _{tot}$ & 6.06 & 6.06 & 6.06 & 6.06 & 6.06 & 6.06 & 6.06 & 6.07 & 
6.08 & 6.09 & 6.1 \\ 
N & 166,802 & 186,689 & 201,779 & 169,888 & 153,551 & 150,941 & 131,385 & 105,659 & 
79150 & 59109 & 47036 \\ \midrule
$R$ & 9.7 & 9.9 & 10.1 & 10.3 & 10.5 & 10.7 & 10.9 & 11.1 & 11.3 & 11.5 & 
11.7 \\ \midrule
$\mathrm{V}_{c}\left( R\right)$ & 228.23 & 228 & 227.38 & 227.05 & 226.3 & 226.22 & 226.55 & 226.4 & 
226.85 & 227.16 & 227.22 \\ 
$\Delta _{stat}$ & 0.11 & 0.12 & 0.12 & 0.12 & 0.13 & 0.13 & 0.14 & 0.15 & 
0.15 & 0.17 & 0.18 \\ 
$\Delta _{tot}$ & 6.11 & 6.12 & 6.12 & 6.12 & 6.13 & 6.13 & 6.14 & 6.15 & 
6.15 & 6.17 & 6.18 \\ 
N & 38,934 & 34,222 & 30,928 & 29,118 & 27,059 & 24,620 & 21,992 & 19,205 & 16,344 & 
13962 & 11977 \\ \midrule
$R$ & 11.9 & 12.1 & 12.3 & 12.5 & 12.7 & 12.9 & 13.1 & 13.3 & 13.5 & 13.7 & 
13.9 \\ \midrule
$\mathrm{V}_{c}\left( R\right)$ & 227.5 & 226.78 & 226.74 & 226.3 & 226.21 & 225.29 & 225.55 & 225.35
& 225.4 & 224.75 & 223.59 \\ 
$\Delta _{stat}$ & 0.19 & 0.22 & 0.23 & 0.26 & 0.29 & 0.34 & 0.39 & 0.45 & 
0.54 & 0.61 & 0.93 \\ 
$\Delta _{tot}$ & 6.19 & 6.22 & 6.23 & 6.26 & 6.29 & 6.34 & 6.39 & 6.45 & 
6.54 & 6.61 & 6.93 \\ 
N & 9787 & 7815 & 6125 & 4860 & 3716 & 2750 & 1993 & 1446 & 965 & 696 & 435\\
\bottomrule
\end{tabular}

\end{table}

\begin{figure}[t]
\includegraphics[height=68mm]{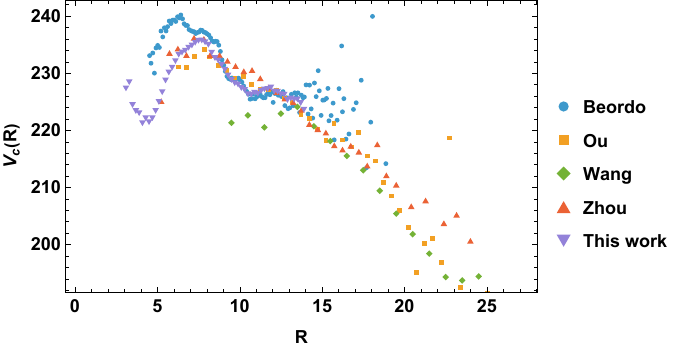}\\ 
\includegraphics[height=70mm, trim=-9mm 0 0 0]{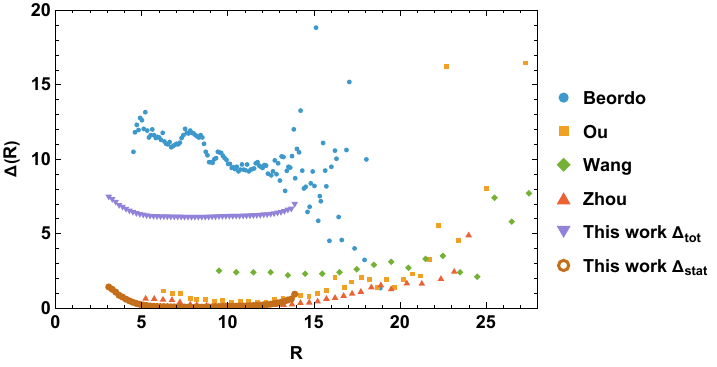}\\%
\caption{Circular velocity curve together with the curves from the studies~\cite{beo,ou,wan22,zho} ({\textbf{top panel}}) 
and corresponding errors ({\textbf{bottom panel}}). Units: $\mathrm{V}_{c}$ [km/s], $R$ [kpc]. }
\label{FGVR}
\end{figure}

In Figure~\ref{FGVR}, we show the circular velocity curve $\mathrm{V}_{c}\left( R\right)$ together with the results of some other studies. For~better clarity, we have illustrated the errors in a separate~part.

\begin{figure}[t]
\includegraphics[width=160mm]{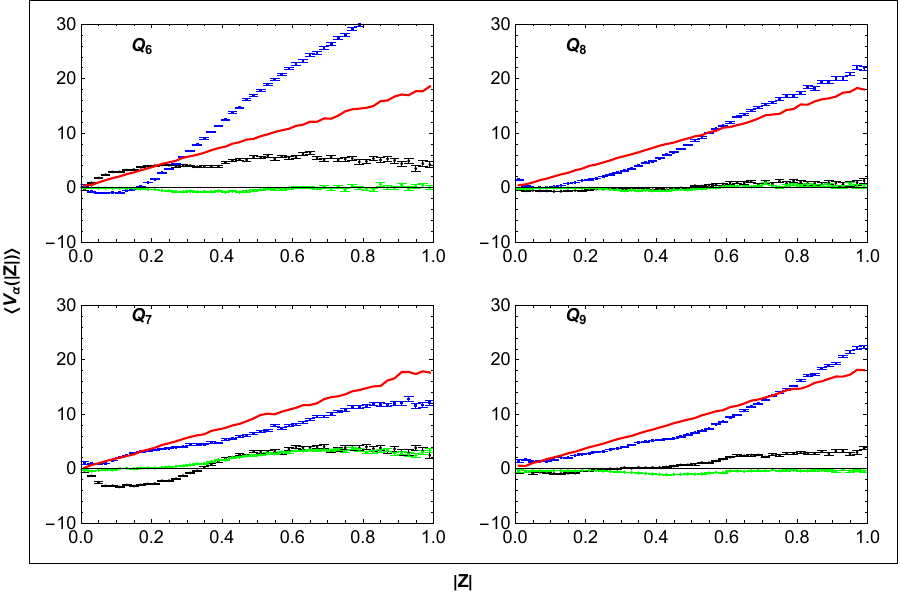} 

\caption{Dependence $\left\langle
V_{\Phi}\right\rangle $ ({blue, simulation red})$,\left\langle
V_{\Theta}\right\rangle $ ({green}), and $\left\langle V_{R}%
\right\rangle $ ({black}) on distance $\left\vert Z\right\vert $ from the galactic plane in sectors Q$_{\alpha}$\textit{. }Units: $V$ [km/s], $Z$ [kpc].}
\label{F2QQV}%
\end{figure}

In Figure~\ref{F2QQV}, we show the velocity curves in the sectors outside the galactic plane as functions of distance $\left\vert Z\right\vert$. Local fluctuations of $\left\langle V_{\Theta}\right\rangle $ and $\left\langle
V_{R}\right\rangle $ in all analysed sectors do not exceed $5$ km/s. On~the other hand, a~clear increase in $\left\langle V_{\Phi}\right\rangle $ is observed in all sectors. The~corresponding simulation, represented by a linear curve controlled by Equation~(\ref{mc1}), which is common to all sectors, is clearly only an approximation. The differences between the measured data and the simulation vary across different sectors. For~$\left\langle V_{\Theta}\right\rangle $ and $\left\langle
V_{R}\right\rangle $, the~simulations yield a net zero (not shown in the figure).

\begin{figure}[t]
\includegraphics[height=70mm]{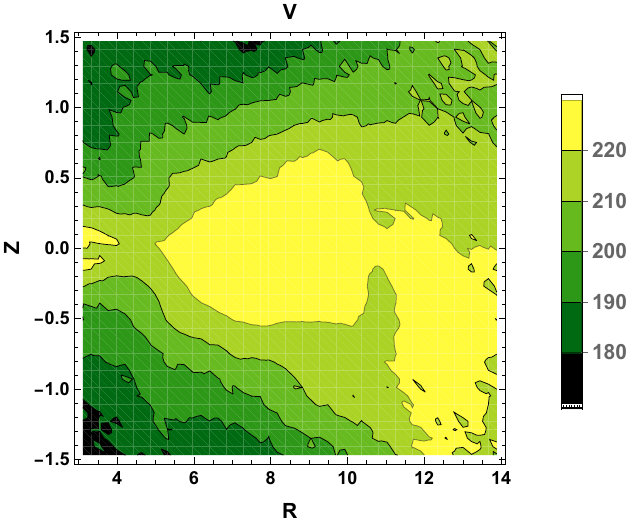}\includegraphics[height=70mm]{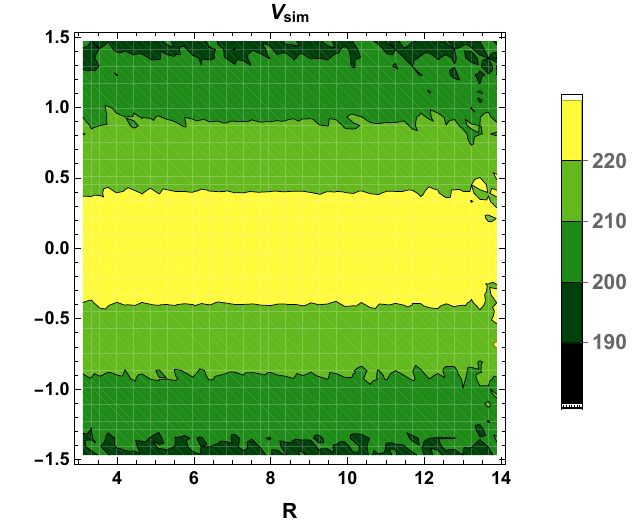}
\ \ \ \includegraphics[height=70mm]{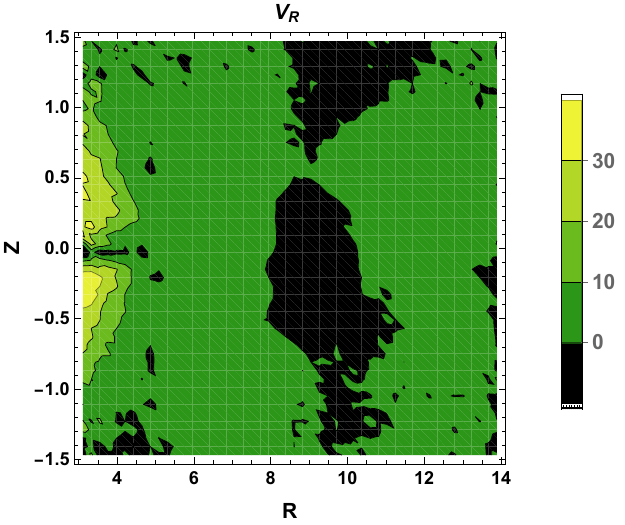}\includegraphics[height=70mm]{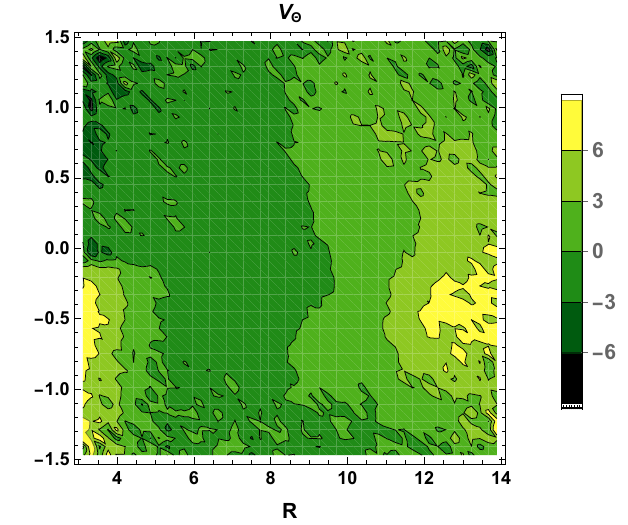}

\caption{{The} 
$R-Z$ diagrams.{ \textbf{Top}:} rotation velocity $\mathrm{V}(R_{c},Z)$ and simulated rotation velocity $\mathrm{V}_{sim}(R_{c},Z)$. {\textbf{Bottom}:} radial velocity $\left\langle V_{R}(R_{c},Z)\right\rangle $ and polar velocity $\left\langle V_{\Theta}(R_{c},Z)\right\rangle $. Units: V [km/s], $R,Z$ [kpc].}%
\label{FRZ}%
\end{figure}

Diagrams of the rotation velocity $\mathrm{V}(R_{c},Z)$, radial velocity $\left\langle V_{R}(R_{c},Z)\right\rangle $, polar velocity $\left\langle
V_{\Theta}(R_{c},Z)\right\rangle $, and simulated rotation velocity $\mathrm{V}_{sim}(R_{c},Z)$ obtained from all sectors are shown in Figure~\ref{FRZ}. 

The plots provide additional information about the kinematics of
the MW in the depicted region. Consistent with the left panel of
Figure~\ref{FGv4}, we again observe that the decrease in rotation velocity
$\mathrm{V}(R_{c},Z)$ in the region $R_{c}\lesssim5$ kpc is accompanied by a
centrifugal component in $\left\langle V_{R}(R_{c},Z)\right\rangle $. We also
observe a slight asymmetry in $\left\langle V_{\Theta}(R_{c},Z)\right\rangle $
\ at smaller $R$. For~$Z<0$, $V_{\Theta}>0$ is dominant; conversely, for~$Z>0$, $V_{\Theta}<0$ dominates.



The rotation velocity $\mathrm{V}(R_{c},Z)$ decreases as $\left\vert
Z\right\vert $ increases. This decrease is more pronounced for smaller $R_{c}$
(sector $Q_{6}$) and less pronounced for larger $R_{c}$ (sector $Q_{7}$). This
pattern is more complex than the pattern from the simulation \textrm{V}%
$_{sim}$(R$_{c}$,Z), which assumes a constant slope $\varkappa$ throughout the
entire region (see Equation~(\ref{mc1})).
\section{Discussion and~Conclusions}
\label{conclusion}
Using a subset of the DR3 data from the Gaia space observatory, which includes information on radial velocities, we studied the 3D kinematics of the MW. This builds on our earlier 2D analysis, which relied on high-statistics data on stars without defined radial velocities. Although~the two approaches differ in both method and input dataset, they yield identical or very close results, for~example, concerning the Sun's local velocity $\Delta\mathbf{V}_{\odot}$ and the mean rotation velocity $\mathrm{V}_{0}$ at radius $R_{\odot}$.

As for $\Delta\mathbf{V}_{\odot}$, the~component perpendicular to the galactic disk plane remains completely unchanged. The~velocity $\mathrm{V}_{0}$ also remains unchanged. It is important to emphasise that both velocities $\Delta\mathbf{V}_{\odot}$ and $\mathrm{V}_{0}$ are obtained completely independently of each other by analysing the stellar velocities in sectors $Q_{1}-Q_{4}$. For~$\mathrm{V}_{0}$ measurements, sectors $Q_{2}$ and $Q_{4}$ are optimal, with~a total sample of 3,737,755 stars. It follows from Equations~\eqref{cor2}--\eqref{cor4} and Table~\ref{Tab3} that the Sun's total velocity  $\mathrm{V}_{0}+\Delta V_{y\odot }=\mathrm{V}_{c}+V_{\odot
}=248.5\pm 5$ km/s (that is, it does not contain the parameter $\mathrm{V}%
_{a}$). Our result is consistent with a very reliable estimate of $251.55\pm
0.98$ km/s based on the longitudinal speed of the radio source Sgr A*\cite%
{reid20} and with the total velocity of $247.4\pm 1.4$ km/s obtained in \cite%
{gra19}. Very good agreement with other measurements also holds for the average rotation velocity $\mathrm{V}_{0}=228\pm 5$ km/s and the Sun's local velocities $(U_{\odot },V_{\odot },W_{\odot })=\left( 10.5\pm 0.7,14.5\pm
0.9,7.33\pm 0.1\right) $ $\mathrm{km/s}$, as~shown in our previous 2D analysis.

Important results are the curves $\left\langle V_{\alpha}(R)\right\rangle
,\mathrm{V}(R),$ and $\left\langle V_{\alpha}(\left\vert Z\right\vert
)\right\rangle $ and the corresponding $R-Z$ diagrams in Figures~\ref{FGv4}, \ref{F2QQV}, and \ref{FRZ}. In~the region $R_{c}\gtrsim5$ kpc in the figures, we observe fluctuations of local kinematic substructures in the range up to $10$ km/s. The~behaviour of the curves in the region $R_{c}\lesssim5$ kpc may be related to the proximity of the galactic~bar.

Our velocity curve $\mathrm{V}_{c}(R)$ in Figure~\ref{FGVR} agrees very well (in the overlap region) with the velocity curves presented in recent studies~\cite{beo,ou,zho,wan22}, which were also obtained using the Gaia DR3 catalogue. Small differences may be due to the parameters for the selection of stars, including the shape of the selection~region.

As can be seen in Figures~\ref{F2QQV} and \ref{FRZ}, the~average rotation velocity of stars slightly decreases with increasing $\left\vert
Z\right\vert $. Therefore, to~determine the curves $\mathrm{V}(R)$ and $\mathrm{V}_{c}(R)$ accurately, the~selection region around the galactic plane should be as narrow as possible. That is why we used a very narrow range $\left\vert b\right\vert <2\deg $ to generate the $\mathrm{V}(R)$ curve. This strict cut affects the curve particularly in the region $R_{c}\lesssim 5$ kpc.

The decrease in rotation velocity with increasing $\left\vert Z\right\vert $ has also been described in~\cite{wan22,ang20}. This effect is consistent with the Besan\c{c}on Galaxy model~\cite{bie18}, which provides a prediction for the region $R_{c}\approx R_{\odot}$. This prediction agrees well with our curves in the sectors $Q_{8}$ and $Q_{9}$ of Figure~\ref{F2QQV}, where $\left\langle
R_{c}\right\rangle \approx R_{\odot}$. Asymmetric drift and a mixed population of thin and thick disks most likely contribute to the negative vertical~gradient.

A useful tool in the analysis of stellar velocities is a simple MC simulation model described in Section~\ref{simul}. The~simulation corresponds to an averaged axisymmetric approximation of the MW, so it may be useful in validating axisymmetric dynamical models or determining the scale of local kinematical substructures out of axial symmetry. The~simulation model describes the distribution of velocities $V_{\Phi}, V_{\Theta}$, and $V_{R}$ in the region $R_{c}\gtrsim5$ kpc. Some discrepancy between the simulated $V_{\Phi
},V_{R}$ and the data in sectors $Q_{1}$ and $Q_{6}$ was mentioned in the previous section (Figures \ref{F1FGc}, \ref{F1RGc}, \ref{F2QQF}, \ref{F2QQR}, and \ref{F2QQV}). The~simulations play an important role in determining the parameter $\mathrm{V}_{0}$ in Figure~\ref{Fgaw}. In~the relevant sectors $Q_{2}$ and $Q_{4}$, the~agreement with the simulation is perfect in the region $r\lesssim3$ kpc. In~these sectors, we also observe very good agreement between the measured data and the simulation for all velocity distributions, as can be seen in Figures~\ref{F1FGc}--\ref{F1RGc}, sector $Q_{5}%
=Q_{2}+Q_{4}$. In~other sectors where simulations yield poorer results, introducing additional free parameters into the MC generator would improve the fit to the~data.

The goal was to present a methodology for the kinematic analysis of the MW. We have shown that our approach makes it possible---using only two external parameters ($R_{\odot }$ and $\mathrm{V}_{a}$) taken from other studies---to derive, with~reasonable accuracy, valuable results that are fully consistent with the findings of other studies. It should be noted that the parameter $\mathrm{V}_{a}$, which characterises asymmetric drift, is a quantity derived from Jeans modelling~\cite{jea1915}, which necessarily assumes an axially symmetric gravitational potential. It would certainly be interesting to attempt to calculate the function $\mathrm{V}%
_{a}(R_{c},Z)$ for subsamples of different ages within the global and, presumably, time-dependent gravitational potential. It is clear, however, that solving such a problem would require further independent~research.

In conclusion, we believe that the proposed methods for calculating the local velocity of the Sun, the~average rotation velocity $V_{0}$, the~velocity curve $\mathrm{V}_{c}(R)$, and, more generally, the~velocity maps $\mathrm{V}(R_{c},Z),$ $\left\langle V_{R}%
(R_{c},Z)\right\rangle $,\ and $\left\langle V_{\Theta}(R_{c},Z)\right\rangle $ represent a useful contribution to the analysis, description, and~understanding of the global kinematics of the MW. The~MC simulation model proved to be an effective tool in our~study.

\vspace{+6pt}

\begin{acknowledgments}%
{This work made use of data from the European Space Agency (ESA) mission
{Gaia} (\url{https://www.cosmos.esa.int/gaia} ({accessed on 8 October 2026})), processed by the 
{Gaia} Data Processing and Analysis Consortium (DPAC,
\url{https://www.cosmos.esa.int/web/gaia/dpac/consortium} ({accessed on 8 October 2026})). {Funding} 
 for the DPAC was provided by national institutions, in~particular the institutions participating in the {Gaia} Multilateral Agreement. We would also like to express our sincere gratitude to the anonymous reviewers for their thorough evaluation of the manuscript and their valuable and constructive suggestions, which have helped us improve its quality.}
\end{acknowledgments}


\appendix
\renewcommand{\theequation}{\thesection.\arabic{equation}}
\renewcommand{\thefigure}{\thesection.\arabic{figure}}
\setcounter{equation}{0}
\setcounter{section}{0}
\setcounter{figure}{0}
\section{Reference~Frames}\label{app11}
\begin{figure}[t]

\hspace{-1em}\includegraphics[width=70mm]{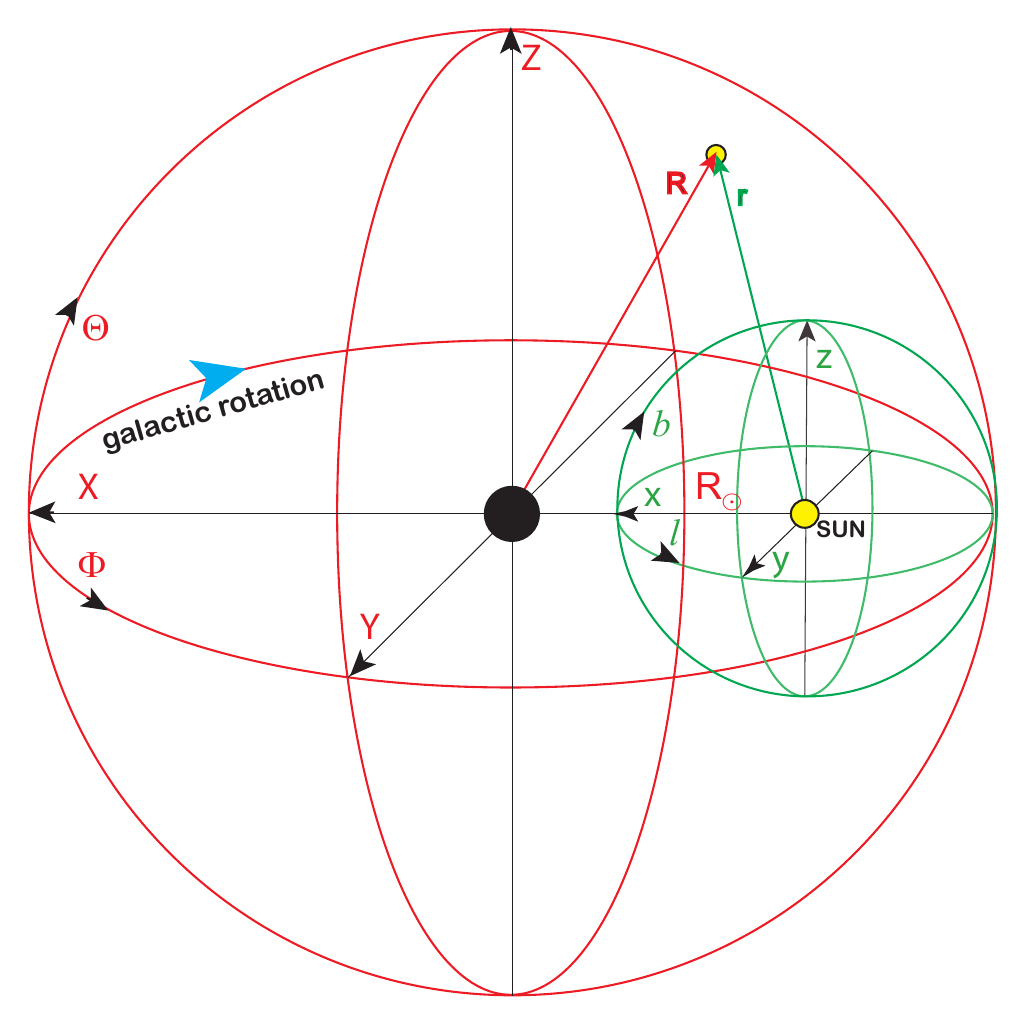}\caption{{Galactic}
 ({green}) and
Galactocentric ({red}) reference frames.}
\label{a1}%
\end{figure}

The relation between galactic and Galactocentric reference frames is
illustrated in Figure~\ref{a1}. 
The axes are defined as\vspace{-6pt}
\begin{align}
x  &  =r\cos l\cos b,\ \ \ \ \ y=r\sin l\cos b,\ \ \ \ \ z=r\sin
b,\label{me2t}\\
X  &  =R\cos\Phi\cos\Theta,\ Y=R\sin\Phi\cos\Theta,\ Z=R\sin\Theta.\nonumber
\end{align}

The $x-$axis therefore points towards the centre of the Galaxy, and the $y-$axis indicates the direction of the MW's rotation. The corresponding coordinates are related as\vspace{-6pt}
\begin{equation}
X=x-R_{\odot};\qquad Y=y;\qquad Z=z. \label{me2u}%
\end{equation}

The radius $R_{\odot}$ represents the distance of the Sun from the galactic centre. The small parameter $Z_{\odot}\approx0.0208$ kpc~\cite{ben19}, defining the position of the Sun above the galactic plane, is neglected in our analysis, as justified in~\cite{za4}. Then, the transformation between the spherical coordinates $\left(  r,l,b\right)  \rightarrow\left(  R,\Phi,\Theta\right)  $ of both frames is defined by the equations
\begin{align}
R  &  =\sqrt{r^{2}+R_{\odot}^{2}-2R_{\odot}x},\qquad\Theta=\arcsin\frac{z}%
{R},\label{me1}\\
\Phi\left(  x>R_{\odot}\right)   &  =\arcsin\frac{y}{R\cos\Theta},\qquad
\Phi\left(  x\leq R_{\odot}\right)  =\pi-\arcsin\frac{y}{R\cos\Theta}\nonumber
\end{align}
and inversely, $\left(  R,\Phi,\Theta\right)  \rightarrow$ $\left(
r,l,b\right)$:
\begin{align}
r  &  =\sqrt{R^{2}+R_{\odot}^{2}+2R_{\odot}X},\qquad b=\arcsin\frac{Z}%
{r}\label{me2}\\
l\left(  X>R_{\odot}\right)   &  =\pi-\arcsin\frac{Y}{r\cos b},\qquad l\left(
X\leq R_{\odot}\right)  =\arcsin\frac{Y}{r\cos b},\nonumber
\end{align}
where $x,y,z,\ X,Y,Z$ are defined by (\ref{me2t}). The spherical coordinates $\left(  R,\Theta\right)  $ are simply related to the cylindrical coordinates $\left(  R_{c},Z\right)  $:\vspace{-6pt}
\begin{equation}
R_{c}=R\cos\Theta;\qquad\ Z=R\sin\Theta. \label{me2a}%
\end{equation}

In our analysis, we need vectors forming the local orthonormal bases in both reference frames:\vspace{-6pt}
\begin{align}
\mathbf{N}_{\Phi}  &  =\left(  -\sin\Phi,\cos\Phi,0\right)  ,\qquad
\mathbf{N}_{\Theta}=\left(  -\cos\Phi\sin\Theta,-\sin\Phi\sin\Theta,\cos
\Theta\right)  ,\label{me3a}\\
\mathbf{N}_{R}  &  =\left(  \cos\Phi\cos\Theta,\sin\Phi\cos\Theta,\sin
\Theta\right)  ,\nonumber\\
\mathbf{n}_{l}  &  =\left(  -\sin l,\cos l,0\right)  ,\qquad\mathbf{n}%
_{b}=\left(  -\cos l\sin b,-\sin l\sin b,\cos b\right)  ,\label{me3b}\\
\mathbf{n}_{r}  &  =\left(  \cos l\cos b,\sin l\cos b,\sin b\right)
,\nonumber
\end{align}
where the vectors $\mathbf{N}_{\alpha}$ and $\mathbf{n}_{\alpha}$ define directions of increasing coordinates $\alpha=\Phi,\Theta,R,\ l,b,r$. For
example, $-\mathbf{N}_{\Phi}$ is the direction of the MW rotation at any point.


\end{document}